\documentclass[useAMS,usegraphicx,usenatbib]{mn2e}
\usepackage{amssymb}
\usepackage{times}
\newcommand{\D}[2]{\frac{\partial #2}{\partial #1}}

\newcommand\bb[1] {\mbox{\boldmath{$#1$}}}
\newcommand\del{\bb{\nabla}}
\newcommand\bcdot{\bb{\cdot}}
\newcommand\btimes{\bb{\times}}
\newcommand\beq{\begin{equation}}
\newcommand\eeq{\end{equation}}
\title[Ambipolar diffusion in the MRI]{Ambipolar diffusion in the magnetorotational instability}
\author[M.W. Kunz \& S.A. Balbus]{Matthew W. Kunz and Steven A. Balbus\thanks{Email: mkunz@alumni.virginia.edu; sb@virginia.edu}\\
Virginia Institute of Theoretical Astronomy, Department of
Astronomy, University of Virginia, PO Box 3818, Charlottesville,
VA 22903, USA}
\date{Released 2003}
\pagerange{\pageref{firstpage}--\pageref{lastpage}} \pubyear{2003}
\def\LaTeX{L\kern-.36em\raise.3ex\hbox{a}\kern-.15em
    T\kern-.1667em\lower.7ex\hbox{E}\kern-.125emX}
\begin{document}
\label{firstpage} \maketitle
\begin{abstract}

The effects of ambipolar diffusion on the linear stability of
weakly ionised accretion discs are examined. Earlier work on this
topic has focused on axial magnetic fields and perturbation
wavenumbers. We consider here more general field and wavenumber
geometries, and find that qualitatively new results are obtained.
Provided a radial wavenumber and azimuthal field are present along
with their axial counterparts, ambipolar diffusion will always be
destabilising, with unstable local modes appearing at well-defined
wavenumber bands. The wavenumber corresponding to the maximum
growth rate need not, in general, lie along the vertical axis.
Growth rates become small relative to the local angular velocity
when the ion-neutral collision time exceeds the orbital time. In
common with Hall electromotive forces, ambipolar diffusion
destabilises both positive and negative angular velocity
gradients. In at least some cases, therefore, uniformly rotating
molecular cloud cores may reflect the marginally stable state of
the ambipolar magnetorotational instability.

\end{abstract}


\section{Introduction}

The behaviour of the magnetorotational instability (MRI) in weakly
ionised gas is a problem of considerable importance, bearing
directly on our understanding of the dynamics of protostellar
discs. The low ionisation regime is characterized not only by
increased Ohmic dissipative losses \citep{j96}, but by Hall
\citep{bate01,war99} and ambipolar diffusion \citep{bla94} effects
as well. The latter two effects are of particular interest, since
they can be destabilising under certain circumstances.

The Hall and ambipolar diffusion regimes are generally important in
distinct regions of parameter space \citep{bate01}: the former is dominant
(along with Ohmic losses) at the high densities typical of inner regions
of protostellar discs, whereas the latter dominates in the outer regions
of such discs (where they are typically observed), and under interstellar
conditions. In this paper, except for a brief excursion in the Appendix,
we confine ourselves to the ambipolar regime, and study the general
axisymmetric behaviour of the MRI in these environments.

The linear and nonlinear effects of ambipolar diffusion on the MRI
have been studied in both the two-fluid \citep{bla94,hs98} and
single-fluid \citep{ml95,war99,sw03} regimes. Our purpose in
revisiting this problem is to highlight a novel result of
practical interest. Previous studies were confined to vertical
wavenumbers only, based on the assumption that this direction
corresponds to the fastest growing local modes. Although this is
true for both ideal and Hall MHD, it turns out that it is not
generally true in the ambipolar limit. Here, the fastest growing
wavenumbers may have both vertical and radial components, and it
is necessary to perform a full axisymmetric analysis to identify
these modes.

The plan of the paper is as follows: In \S 2, we discuss the basic
formulation of the problem, including a presentation of temperature and
number density regimes associated with non-ideal MHD effects. Section 3 is
a linear analysis of differentially rotating discs for axial wavenumber
and field configurations. This serves to unite this paper with previous
work on ambipolar diffusion. Section 4 extends this analysis for
arbitrary field geometries and perturbation wavenumbers, and contains
the key results of the paper.  In \S 5, we present a brief summary.

\section{Preliminaries}

The fundamental fluid equations are mass conservation,
\begin{equation}
\D{t}{\rho} + \del\bcdot(\rho\bb{v}) = 0,
\end{equation}
and the equation of motion,
\begin{eqnarray}
\lefteqn{\rho\D{t}{\bb{v}} +
\left(\rho\bb{v}\bcdot\del\right)\bb{v} = -\del\left(P +
\frac{B^{2}}{8\pi}\right) - \rho\del\Phi +
\left(\frac{\bb{B}}{4\pi}\bcdot\del\right)\bb{B}.}\nonumber\\*
\end{eqnarray}
Our notation is standard: $\rho$ is the mass density, $\bb{v}$ the
velocity, $P$ the gas pressure, $\bb{B}$ the magnetic field, and
$\Phi$ the gravitational potential. The density, velocity, and
pressure all refer to the dominant neutral species.

The induction equation is (see e.g., Balbus \& Terquem 2001):
\begin{equation}\label{induc}
\D{t}{\bb{B}} = \del\btimes\left[\bb{v}\btimes\bb{B} -
\frac{4\pi\eta\bb{J}}{c} - \frac{\bb{J}\btimes\bb{B}}{n_{e} e} +
\frac{(\bb{J}\btimes\bb{B})\btimes\bb{B}}{c\gamma\rho_{i}\rho}\right].
\end{equation}
Here $\eta$ is the resistivity, $c$ the speed of light, $\bb{J}$
the current density, $n_e$ the electron density, $e$ the magnitude
of the electron charge, $\gamma$ the electron-ion drag
coefficient, and $\rho_i$ the ion mass density. Numerical values
of $\eta$ and $\gamma$ are \citep{dra83,bate01}:
$$
\eta = 234 \left(n\over n_e\right) T^{1/2} {\rm cm}^2 {\rm
s}^{-1}, \quad  \gamma = 2.75\times 10^{13} {\rm cm}^3\,{\rm
s}^{-1}\, {\rm g}^{-1},
$$
where $n$ is the number density of the neutrals and $T$ is the
temperature. Finally, for future reference,
we define the Alfv\'en velocity in the usual
way:
\begin{equation}
\bb{v_{\rm A}} = { \bb{B}\over \sqrt{4\pi\rho}}.
\end{equation}

The terms on the right side of equation (\ref{induc}) correspond
respectively to Faraday induction, Ohmic resistivity, Hall
electromotive forces (HEMFs), and ambipolar diffusion. The
relative ratio of the ambipolar to Hall terms is given by
\citep{bate01}:
\begin{equation}
{A\over H} \sim \left(\frac{ 10^{13}}{n}\right)^{1/2}
\left(\frac{T}{10^{3}}\right)^{1/2}\left(\frac{v_{\rm
A}}{c_s}\right),
\end{equation}
where $c_s$ is the isothermal sound speed.  Assuming that the
final two factors are each about 0.1, we see that a neutral
density below about $10^9$ cm$^{-3}$ brings us into the ambipolar
diffusion regime (see Fig. 1), and we shall assume that this
restriction is satisfied. We retain the resistivity in \S 3,
however, to illustrate the simple relation of ambipolar diffusion
to Ohmic dissipation for the case of axial field geometry,
and in the Appendix, where both Hall and Ohmic terms
are included in the derivation of a very general dispersion
relation.


\begin{figure}
\includegraphics[width=84mm]{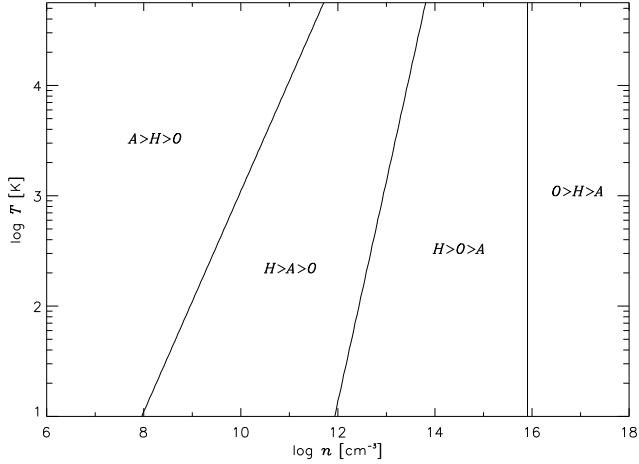}
\caption{Regions of ambipolar ($A$), Hall ($H$), and Ohmic ($O$)
dominance. Numbers are taken from eqs.\ (24) and (25) from
\citet{bate01}, assuming $v_{\rm A}/c_s=0.1$. Note that there is
no explicit dependence on radius, central mass, or ionisation
fraction. The region of ambipolar dominance of interest is the
left-most wedge.}
\end{figure}

\section{Axial Fields and Wavenumbers}

\subsection{Stability}

We consider first the local stability of a differentially rotating
disc threaded by a weak vertical field, $\bb{B} = B\bb{e_{Z}}$, in
the presence of ambipolar diffusion. We use standard cylindrical
coordinates $(R,\phi,z)$ with the origin at the disc centre, and
consider plane wave perturbations that depend only on $z$.  Linearized
quantities (indicated by $\delta$ notation) are proportional to
$\exp (\sigma t - ikz)$, where $k$ is the vertical wave number. In
the Boussinesq limit, this corresponds to fluid displacements in the
plane of the disc, so vertical structure is unimportant. Under these
circumstances, pressure, density, vertical velocity, and vertical magnetic
field perturbations all vanish. The two-fluid version of this problem
has been considered by \citet{bla94}; we shall compare the results of
our single-fluid treatment with those of these authors.

Stability is most easily assessed by working in the limit
$\sigma\rightarrow 0$, assuming that the stable-unstable
transition is bridged by a $\sigma=0$ solution. The linearized
equations of motion are
\begin{equation}
-2\Omega\delta v_{\phi} - \frac{ikB}{4\pi\rho}\delta B_{R} = 0,
\end{equation}\begin{equation}
\frac{\kappa^{2}}{2\Omega}\delta v_{R} -
\frac{ikB}{4\pi\rho}\delta B_{\phi} = 0,
\end{equation}
where $\kappa$ is the epicyclic frequency defined by
\begin{equation}
\kappa^{2} = 4\Omega^{2} + \frac{d\Omega^2}{d\ln R}.
\end{equation}
The linearized induction equations are
\begin{equation}
k^{2}\eta'\delta B_{R} - ikB\delta v_{r} = 0,
\end{equation}\begin{equation}
k^{2}\eta'\delta B_{\phi} - \frac{d\Omega}{d\ln R}\delta B_{R} -
ikB\delta v_{\phi} = 0,
\end{equation}
where
\begin{equation}
\eta' \equiv \eta + \frac{v^{2}_{\rm A}}{\gamma\rho_{i}}
\end{equation}
is the Ohmic resistivity modified by ambipolar diffusion.

Solving the above system of equations, we find
\begin{equation}
\kappa^{2}k^{2}\eta'^{2} + v^{2}_{\rm
A}\left(\frac{d\Omega^{2}}{d\ln R} + k^{2} v^{2}_{\rm A}\right) =
0.
\end{equation}
This is the desired marginal stability condition. It is easily
shown that the left hand side should be negative for instability.
The {\it instability} criterion is then,
\begin{equation}
k^{2}v^{2}_{\rm A} < -\frac{d\Omega^{2}}{d\ln R}
\left[1+\frac{\kappa^{2}\eta'^{2}}{v^{4}_{\rm A}}\right]^{-1},
\end{equation}
a more restrictive condition than the classical MRI.

It is often the case that Ohmic resistivity is small compared with
ambipolar diffusion, which was the limit assumed by \citet{bla94}.
If we take $\eta \rightarrow 0$, $\eta ' \rightarrow v^2_{\rm
A}/\gamma\rho_i$, the instability criterion becomes
\begin{equation}\label{instab1}
k^2 v^2_{{\rm A}i} < -\frac{d\Omega^2}{d\ln
R}\left(\frac{\gamma^2\rho_i\rho}
{\kappa^2+\gamma^2\rho^2_i}\right),
\end{equation}
where we have introduced the {\em ion} Alfv\'{e}n velocity
\begin{equation}
v_{{\rm A}i}\equiv v_{\rm
A}\left(\frac{\rho}{\rho_i}\right)^{1/2}.
\end{equation}
The analogous two-fluid instability criterion obtained by
\citet{bla94} was
\begin{eqnarray}\label{instabbb94}
\lefteqn{k^2 v^2_{{\rm A}i} < -{d\Omega^2\over d\ln R}
\left({\kappa^2 + \gamma^2\rho_i\rho} \over
{\kappa^2+\gamma^2\rho^2_i}\right) \quad {\rm (BB94\
instability\ criterion).}}\nonumber\\* \mbox{}
\end{eqnarray}
Clearly, the inequality (\ref{instab1}) is consistent with the
inequality (\ref{instabbb94}) provided that
$$
\gamma\rho \gg \kappa,
$$
a quantitative implementation of the assumption of negligible ion
inertia.

\subsection{Axial Wavenumber Dispersion Relation}

We next obtain the full dispersion relation for finite $\sigma$.
The linearized radial and azimuthal equations of motion are now
\begin{equation}
\sigma\delta v_{R} - 2\Omega\delta v_{\phi} -
\frac{ikB}{4\pi\rho}\delta B_{R} = 0,
\end{equation}\begin{equation}
\sigma\delta v_{\phi} + \frac{\kappa^{2}}{2\Omega}\delta v_{R} -
\frac{ikB}{4\pi\rho}\delta B_{\phi} = 0.
\end{equation}
The same components of the linearized induction equation are
\begin{equation}
(\sigma + k^{2}\eta')\delta B_{R} - ikB\delta v_{r} = 0,
\end{equation}\begin{equation}
(\sigma + k^{2}\eta')\delta B_{\phi} - \frac{d\Omega}{d\ln
R}\delta B_{R} - ikB\delta v_{\phi} = 0.
\end{equation}
The resulting dispersion relation is
\begin{equation} \sigma^{4} +
2k^{2}\eta'\sigma^{3} + \mathcal{B}_{2}\sigma^{2} +
2k^{2}\eta'(\kappa^{2} + k^{2}v^{2}_{\rm A})\sigma +
\mathcal{B}_{0} = 0,
\end{equation}
where the constants $\mathcal{B}_{2}$ and $\mathcal{B}_{0}$ are
given by
\begin{equation}
\mathcal{B}_{2} = \left[\kappa^{2} + 2k^{2}v^{2}_{\rm A} +
k^{4}\eta'^{2} + \frac{d\Omega^{2}}{d\ln R}\right],
\end{equation}\begin{equation}
\mathcal{B}_{0} = k^{2}v^{2}_{\rm A}\left[\frac{d\Omega^{2}}{d\ln
R} + k^{2}v^{2}_{\rm A}\right] + \kappa^{2}k^{4}\eta'^{2}.
\end{equation}
Ambipolar diffusion acts on axial wavenumber disturbances only as
though it were a field-dependent additive resistivity, $\eta'$.
Adding other field components while retaining $\bb{k}=k\bb{e_Z}$,
or adding other wavenumber components while retaining
$\bb{B}=B\bb{e_Z}$, does not change the dispersion relation.
Qualitative changes are introduced only if nonaxial wavenumbers as
well as nonaxial field components are considered.

\section{General Analysis}
\subsection{Preliminary Assumptions}
We now consider the axisymmetric behaviour of the instability with
more general field geometries and wavenumbers. We ignore buoyancy,
so our analysis holds either either for a barotropic disc, or
locally at the midplane. The perturbation wavevector is
\begin{equation}
\bb{k} = k_{R}\bb{e_{R}} + k_{Z}\bb{e_{Z}},
\end{equation}
and disturbances have space-time dependence
exp$(i\bb{k}\bcdot\bb{r} + \sigma t)$. In the presence of shear, a
toroidal magnetic field $B_\phi$ would grow linearly with time if
a radial field component $B_R$ were also present:
\begin{equation}
B_{\phi}(t) = B_{\phi}(0) + tB_{R}\frac{d\Omega}{d\ln R}.
\end{equation}
(This ignores the effect of ambipolar diffusion in the unperturbed
background disc.) Since $B_{\phi}$ explicitly enters the
dispersion relation and we are assuming a simple exponential time
behaviour for the perturbations, we set $B_{R}=0$ in the analysis
for self-consistency. The inclusion of a $B_R$ component should
not qualitatively affect the stability behaviour, though it may
affect growth rates. We shall also ignore Ohmic losses, as
ambipolar diffusion generally dominates over this term (cf.\ Fig.\
1).

\subsection{Dispersion Relation}

The linearized mass conservation equation in the Boussinesq limit
is:
\begin{equation}
k_{R}\delta v_{R} + k_{Z}\delta v_{Z} = 0.
\end{equation}
Using this to
eliminate $\delta v_Z$ from the equations of motion leads to the
radial and azimuthal forms
\begin{equation}\label{dyn1}
\sigma \delta v_R - {k^2_Z\over k^2} 2 \Omega \delta v_\phi +
i\frac{\bb{k}\bcdot\bb{B}}{4\pi\rho} \delta B_R = 0,
\end{equation}
\begin{equation}\label{dyn2}
\sigma \delta v_\phi + {\kappa^2\over 2\Omega}\delta v_R - i
\frac{\bb{k}\bcdot\bb{B}}{4\pi\rho}\delta B_\phi = 0.
\end{equation}
By similarly eliminating $\delta B_Z$ via a vanishing divergence
condition, only the $R$- and $\phi$-components of the induction
equation are needed. The linearized induction equations are
\begin{eqnarray}
\lefteqn{ \left(\sigma + {k^{2}B^2_Z\over 4\pi \gamma
\rho_i\rho}\right) \delta B_R
-i (\bb{k}\bcdot\bb{B})\delta v_R} \nonumber\\
& & - {k_R B_\phi(\bb{k}\bcdot\bb{B})\over 4\pi \gamma \rho_i
\rho}\, \delta B_\phi = 0,
\end{eqnarray}

\begin{eqnarray}
\lefteqn{ \left(\sigma + {k^{2}B_\phi^2 +
(\bb{k}\bcdot\bb{B})^2\over 4\pi \gamma \rho_i\rho}\right) \delta
B_\phi
-i (\bb{k}\bcdot\bb{B})\delta v_\phi} \nonumber \\
& & - \left[ {d\Omega\over d\ln R} +{k^2\over k^2_Z}{k_R
B_\phi(\bb{k}\bcdot\bb{B})\over 4\pi\gamma\rho_i \rho}\right]
\delta B_R = 0.
\end{eqnarray}
The resulting dispersion relation is
\begin{equation}
\sigma^{4} + \mathcal{C}_3\sigma^{3} + \mathcal{C}_{2}\sigma^{2} +
\mathcal{C}_1\sigma + \mathcal{C}_0 =0,
\end{equation}
with
\begin{equation}
\mathcal{C}_3 = {k^2B^2 + (\bb{k}\bcdot\bb{B})^2\over
4\pi\gamma\rho_i \rho}
\end{equation}
\begin{eqnarray}
\lefteqn{ \mathcal{C}_2 = \frac{k_Z^2}{k^2}\kappa^2 + 2
(\bb{k}\bcdot\bb{v_{\rm A}})^2 + \left[
\frac{kB(\bb{k}\bcdot\bb{B})}{4\pi\gamma\rho_i\rho}\right]^2} \nonumber\\ &&
 - {k_RB_\phi(\bb{k}\bcdot\bb{B}) \over 4\pi\gamma \rho_i \rho}\,
{d\Omega\over d\ln R}
\end{eqnarray}
\beq \mathcal{C}_1 = \left( (\bb{k}\bcdot\bb{v_{\rm A}})^2 +
{k_Z^2\over k^2}\kappa^2 \right) \mathcal{C}_3 \eeq
\begin{eqnarray}\label{c0}
\lefteqn{ \mathcal{C}_0 = (\bb{k}\bcdot\bb{v_{\rm
A}})^2\left[(\bb{k}\bcdot\bb{v_{\rm A}})^2 +
\frac{k^2_Z}{k^2}\frac{d\Omega^2}{d\ln R}\right] +
\frac{k_Z^2}{k^2}\kappa^2\left[
\frac{kB(\bb{k}\bcdot\bb{B})}{4\pi\gamma\rho_i\rho}\right]^2}\nonumber\\
&& -\left( (\bb{k}\bcdot\bb{v_{\rm A}})^2 +
\frac{k_Z^2}{k^2}\kappa^2 \right) { k_R
B_\phi(\bb{k}\bcdot\bb{B})\over 4\pi\gamma\rho_i \rho}\,
{d\Omega\over d\ln R}.
\end{eqnarray}

\subsection{Stability}

We assume that the transition from stability to instability
proceeds through $\sigma = 0$, a point that can be confirmed
numerically. In the limit $\sigma \rightarrow 0$,
\begin{equation}
\sigma = - {\mathcal{C}_0 / \mathcal{C}_1}.
\end{equation}
With $\mathcal{C}_1> 0$, the
instability requirement is then
$$
\mathcal{C}_0 < 0,
$$
From equation (\ref{c0}), it is apparent that $\mathcal{C}_0$
consists of the usual MRI terms plus explicit ambipolar diffusion
terms. The first of these is simply an effective resistivity, but
the second introduces a novel form of coupling: in formal
Cartesian index notation $(i,j,k)$ it is
$$
k_i B_j{\partial v_j/\partial x_i}.
$$
It is the dominant ambipolar term in $\mathcal{C}_0$ when
$\gamma\rho_i$ is large, and can always be chosen to be
destabilising. Since this term is nonvanishing only in the
presence of a nonaxial wavenumber component, it is easy to show
that the fastest growing wavenumber need not, in general, lie
along the $z$-axis.

This has important consequences. Let us view the marginal
stability transition $\mathcal{C}_0 = 0$ as an equation for
$$
X \equiv \frac{(\bb{k}\bcdot\bb{v_{\rm A}})^2}{\Omega^2}
$$
as a function of $k_R/k_Z$, for a given set of disc parameters.
Transition to instability is present only if there are solutions
to this equation with $X$ real and positive. With
$$
\tilde{\kappa}^2=\kappa^2/\Omega^2,
$$
one finds that the $\mathcal{C}_0=0$ requirement is satisfied when
\begin{eqnarray}\label{Xinstab}
\lefteqn{X = -\frac{k^2_Z}{k^2}\frac{d\ln\Omega} {d\ln R} \left[
{\displaystyle{ {2-
\tilde{\kappa}^2\frac{\Omega}{\gamma\rho_i}\frac{k_R}{k_Z}\frac{B_\phi}{B_Z}}
}} \over{\displaystyle{
1+\frac{\Omega}{\gamma\rho_i}\left(\tilde{\kappa}^2
\frac{\Omega}{\gamma\rho_i}\frac{B^2}{B^2_Z}-
\frac{k_R}{k_Z}\frac{B_\phi}{B_Z}\frac{d\ln\Omega}{d\ln
R}\right)}}\right].}\nonumber\\*&&\mbox{}
\end{eqnarray}
The denominator vanishes at a wavenumber ratio
\begin{equation}
{k_R\over  k_Z} = \left( {\Omega\over\gamma\rho_i} { B_\phi\over B_Z}
{d\ln\Omega\over d\ln R} \right)^{-1} \left[ 1 + \left(\tilde{\kappa}
{B\over B_Z}{ \Omega\over\gamma\rho_i}\right)^2 \right].
\end{equation}
It is straightforward to show that there will always be solutions
with large positive values of $X$ near this selected value of
$k_R/k_Z$. These large wavenumber modes are well localised in a
WKB sense. In other words, {\em ambipolar diffusion always
destabilises differential rotation,} whether the profile is
increasing inward or outward. Despite the presence of a resistive
term in the dispersion relation, there are unstable modes at large
nonaxial wavenumbers. Moreover, because fluid motions do not
simply bend field lines locally when ambipolar effects dominate
(loss of flux-freezing), some of the stabilising effects of large
wavenumber magnetic tension seen in the standard MRI are lost.

The modification of the MRI in the presence of ambipolar diffusion
does not change the fundamental cause of the instability, which is
based on rotational dynamics: angular momentum is removed from
fluid elements with less angular momentum and given to fluid
elements with more angular momentum, an intrinsically
destabilising process. What is complicated by the presence of
ambipolar diffusion is the fluid element ``tethering,'' now no
longer a simple, spring-like magnetic tension based on
flux-freezing. The radial component of the disturbed magnetic
field, for example, depends both upon azimuthal as well as radial
motions, and field lines do not follow fluid elements. One
consequence is the destabilisation of outwardly increasing
rotation profiles. In astrophysical settings, the primary interest
is not outwardly increasing angular velocity profiles of course,
but the additional destabilisation that attends the usual
outwardly decreasing profiles. Ambipolar diffusion does not,
however, lead to growth rates in excess of $0.75\Omega$
\citep{bal92}. Indeed, as shown in the following section, growth
rates become a small fraction of $\Omega$ when the ambipolar
parameter $\Omega/\gamma\rho_i$ significantly exceeds unity.

\begin{figure}
\includegraphics[width=84mm]{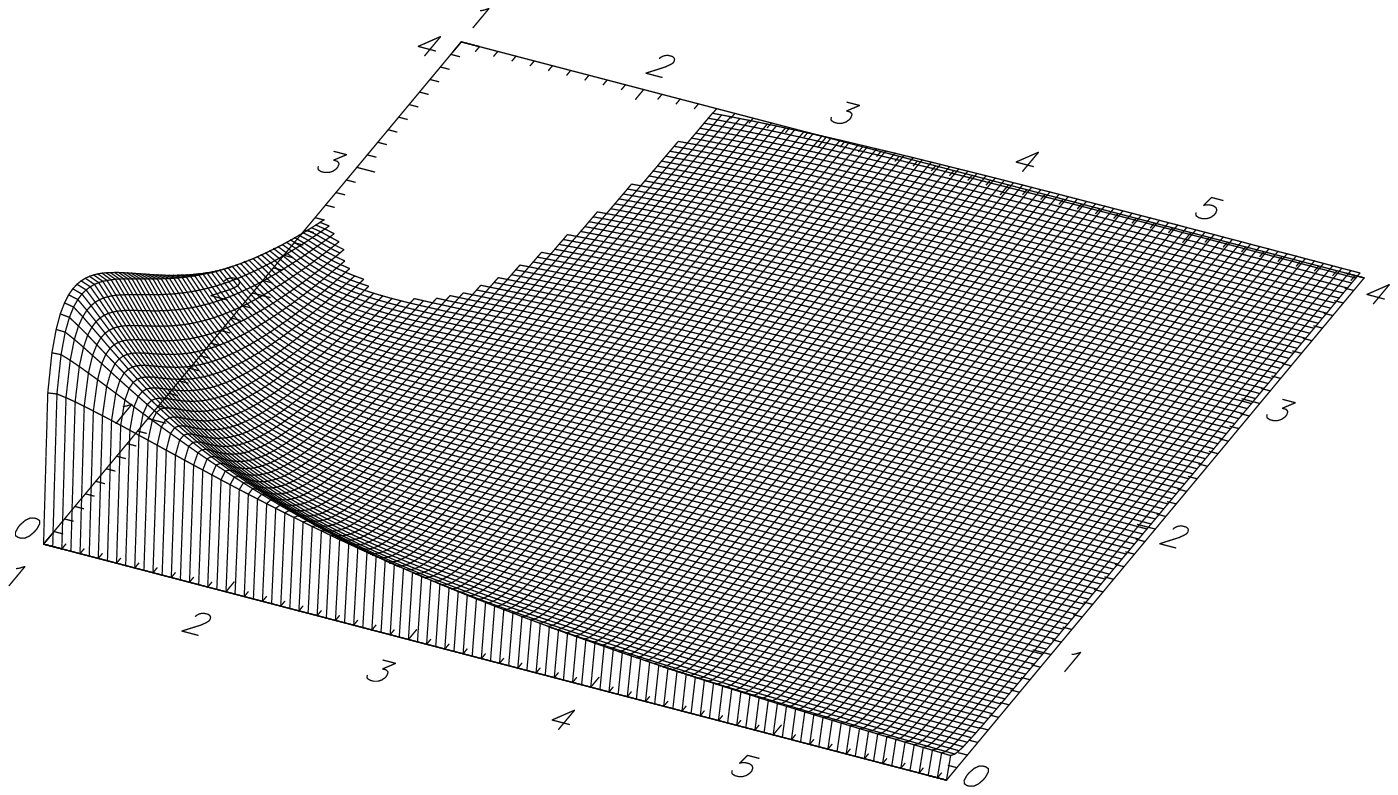}
\includegraphics[width=84mm]{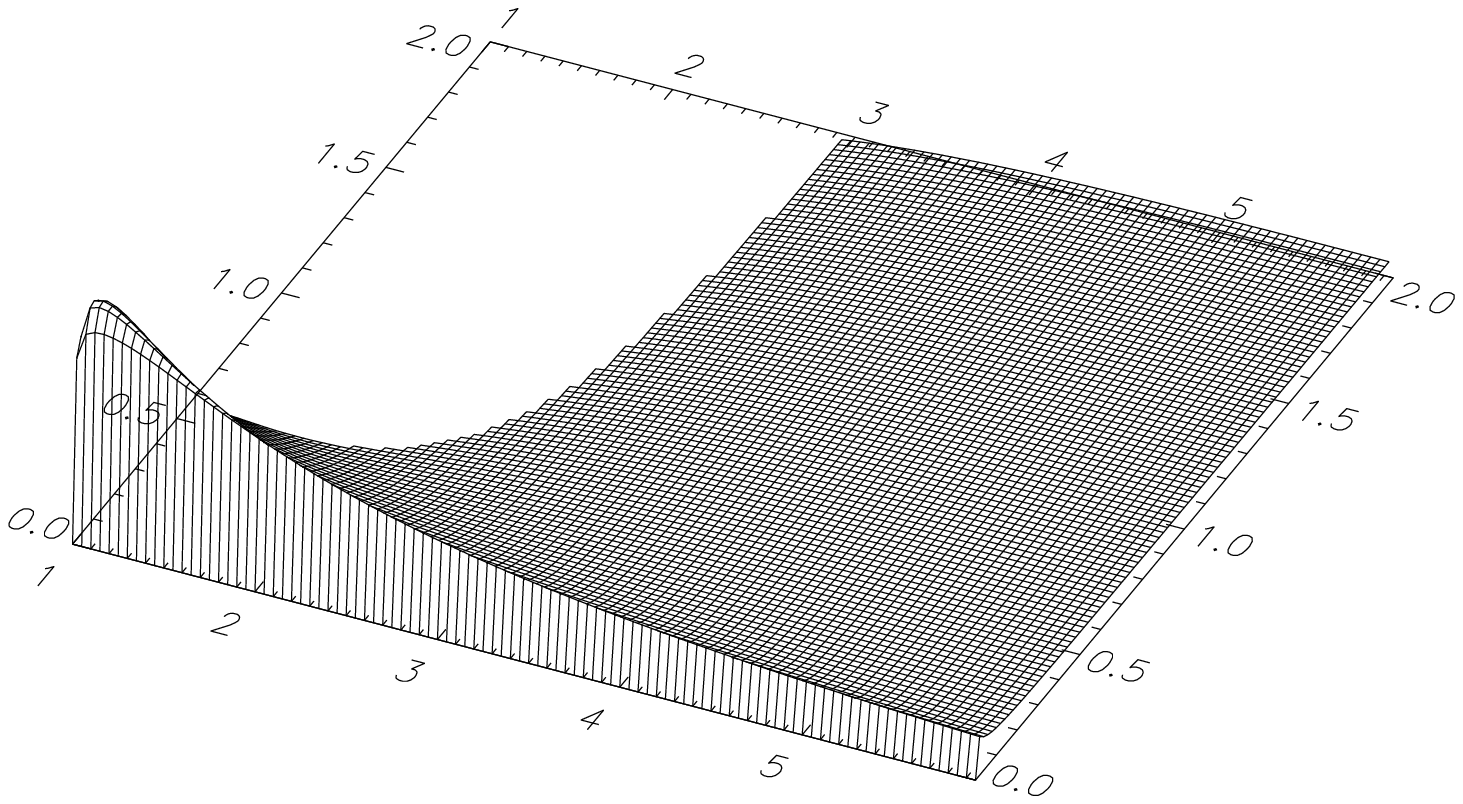}
\includegraphics[width=84mm]{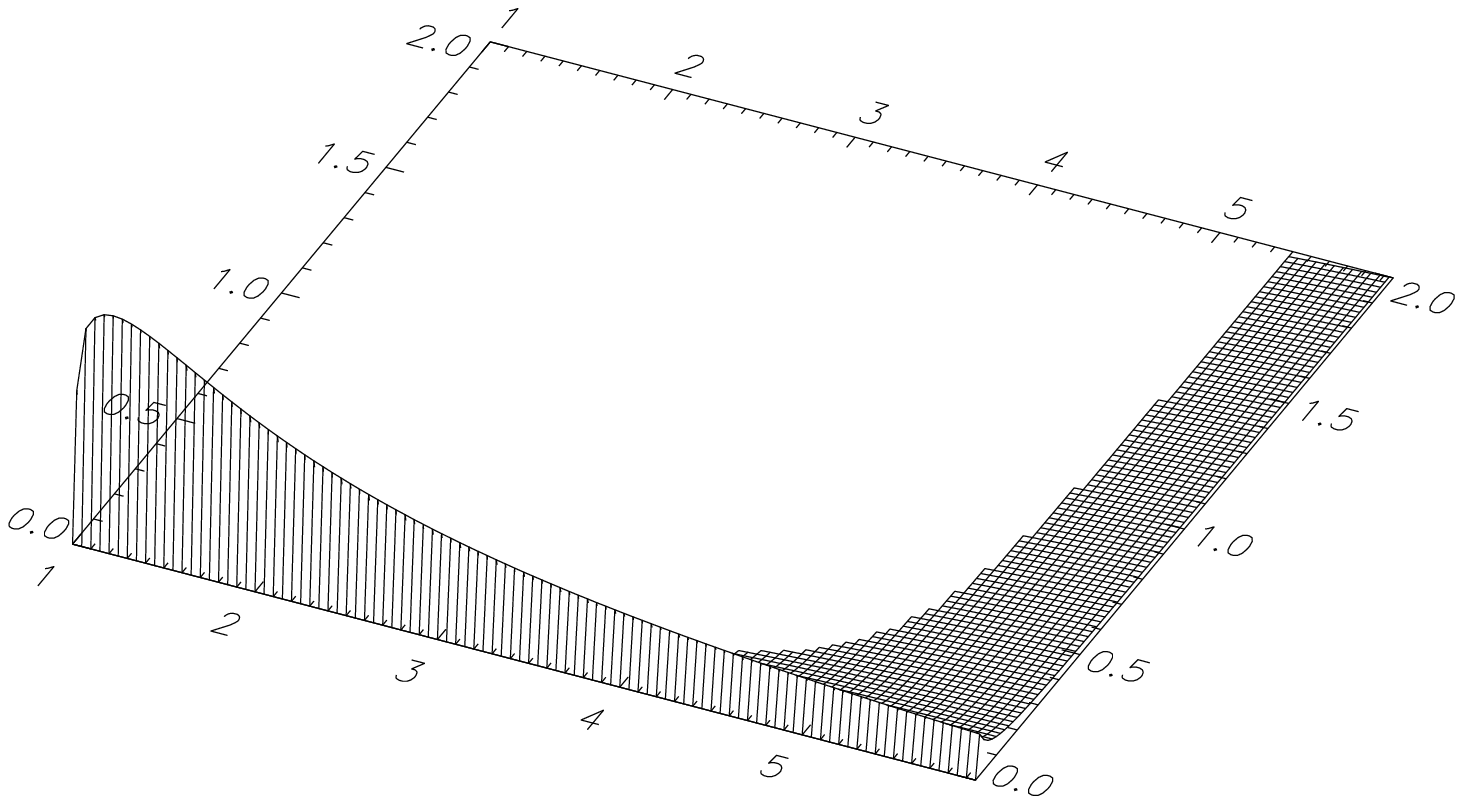}
\caption{Instability growth rates in the $(\bb{k}\bcdot\bb{v_{\rm
A}})^2/\Omega^2$ (ordinate), $|k/k_Z|$ (abscissa) plane for a
Keplerian disc with $B_\phi/B_Z=5$ and $\Omega/\gamma\rho_i=0.1,$
0.5, and 1 (from top to bottom). Only regions of instability are
shown, with the height proportional to the growth rate. The
maximum growth rates, from top to bottom respectively, are
$0.41\Omega$, $0.17\Omega$, and $0.11\Omega$. Note that the
classical MRI maximum growth rate $0.75\Omega$ occurs at
$(\bb{k}\bcdot\bb{v_{\rm A}})^2/\Omega^2 =15/16$ and $k/k_Z =1$.}
\end{figure}

\begin{figure}
\includegraphics[width=84mm]{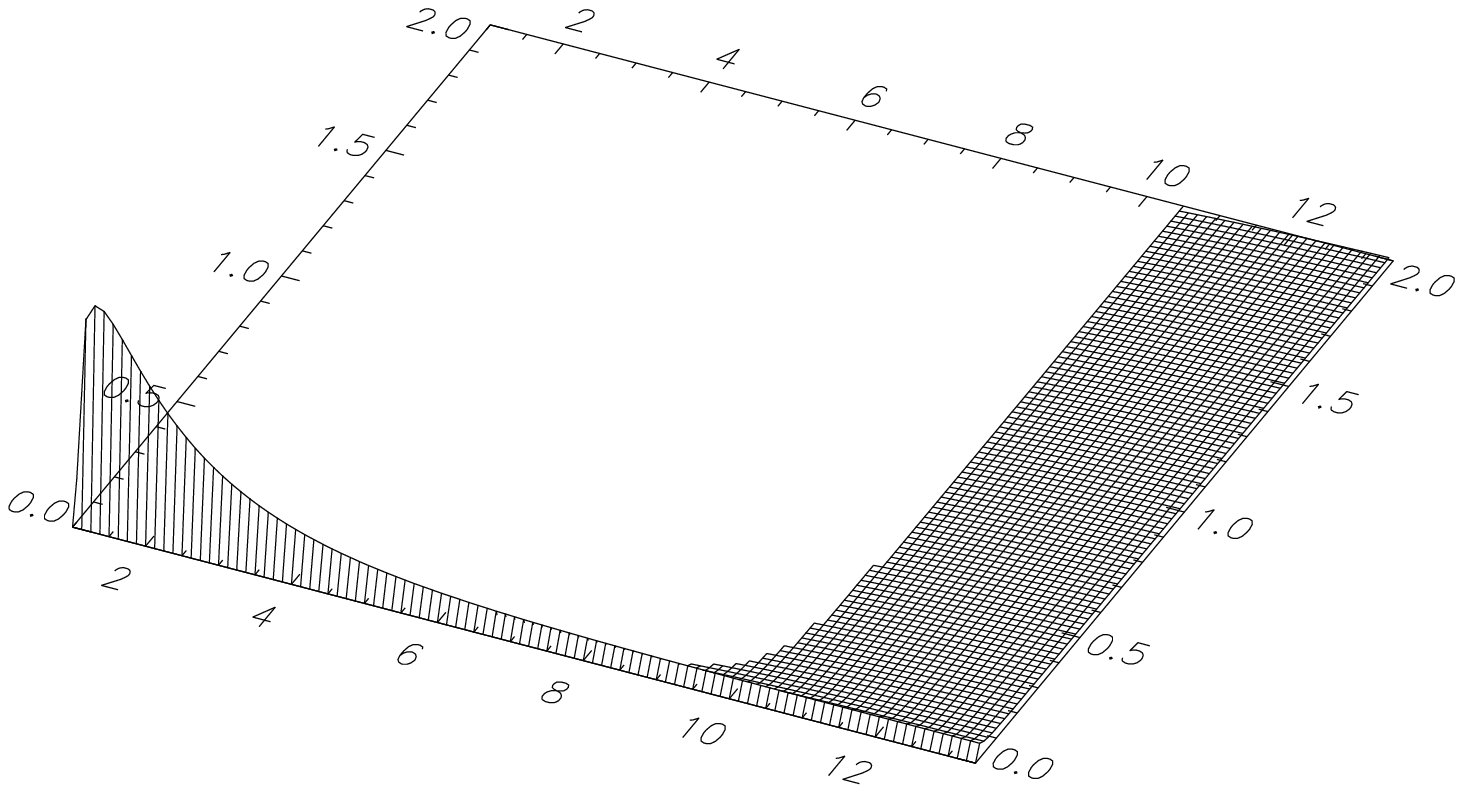}
\includegraphics[width=84mm]{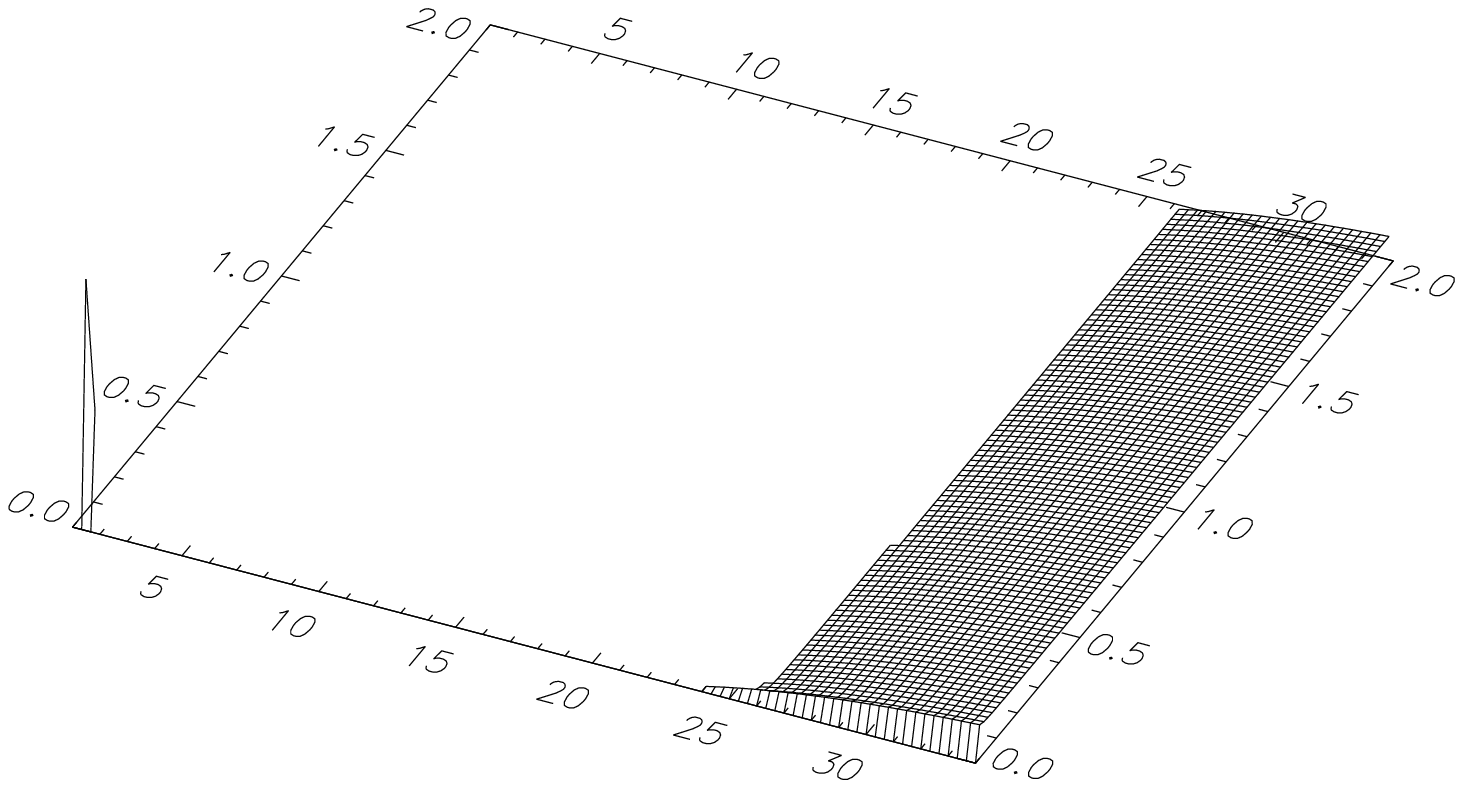}
\includegraphics[width=84mm]{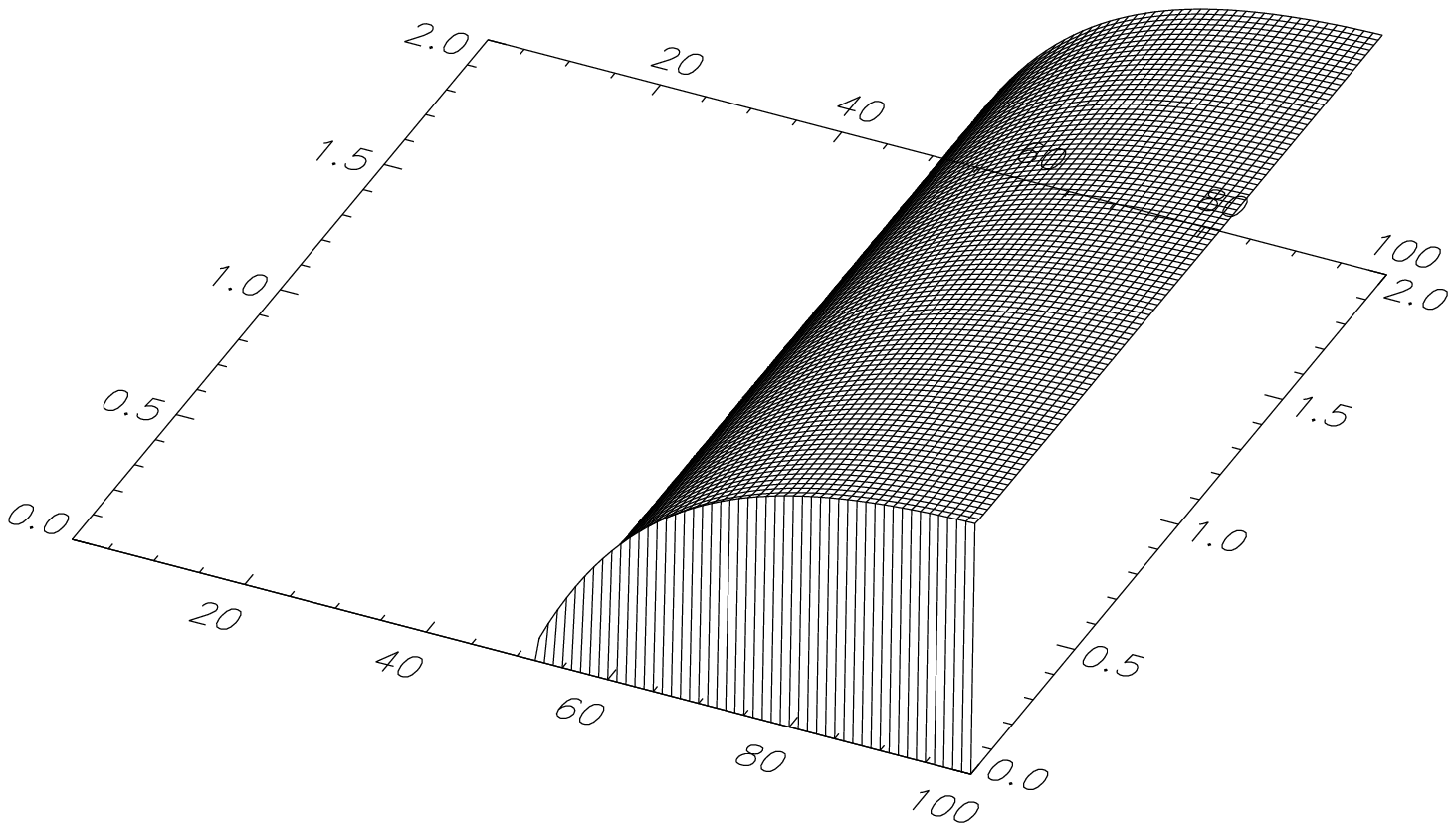}
\caption{As in Fig.\ 2, but for $\Omega/\gamma\rho_i=2,$ 5, and 10
(from top to bottom). The maximum growth rates, top to bottom
respectively, are $0.08\Omega$, $0.01\Omega$, and $9\times
10^{-4}\Omega$.}
\end{figure}

\subsection{Growth Rates}

It is of interest to calculate explicitly some representative growth rates
in the two-dimensional phase space described by the wavenumber parameters
$(\bb{k}\bcdot\bb{v_{\rm A}})/\Omega$ and $k/k_Z$. In Figure 2, we present
three-dimensional plots of the growth rate for representative cases with
$\Omega/\gamma\rho_i \le 1$. While the general tendency of ambipolar
diffusion is to lower the maximum growth rate below $0.75\Omega$, there
is also a significant extension of the domain of unstable non-axial
wavenumbers.

As $\Omega/\gamma\rho_i$ approaches and exceeds unity, the most
unstable modes are associated with a near cancellation of the
resistive and nonresistive (destabilising) ambipolar terms in the
$\mathcal{C}_0$ term of the dispersion relation. The modes appear
at progressively larger values of $k_R/k_Z$. The maximum growth
rate decreases sharply from $\sim 0.1\Omega$ at $\Omega \sim
\gamma\rho_i$ to $\sim 0.001\Omega$ at $\Omega \sim
10\gamma\rho_i$ (see Fig. 3). Once the rotation frequency drops
below the ion-neutral collision frequency, the growth time becomes
very long compared with an orbital time. The instability remains
viable, however, if its growth rate remains shorter than other
evolutionary timescales.

\section{Summary}

In this paper we have examined the axisymmetric magnetorotational
instability in the presence of ambipolar diffusion. Our results
should find their primary applicability in molecular discs on
interstellar or galactic scales, as well as the low density
outermost regions of protostellar discs. We have allowed radial
and axial wavenumber geometries, and azimuthal and axial magnetic
field components. The inclusion of a radial magnetic field
component would induce a linear time dependence in the azimuthal
field component. Since $B_\phi$ appears explicitly in the
dispersion relation, we considered only the case of a vanishing
background radial field component.

The combination of nonaxial wavenumbers and magnetic fields
greatly enhances the response of a differentially rotating gas to
ambipolar diffusion, compared with the axial geometries that are
usually analysed. In particular, the wavevector corresponding to
the maximum growth rate is nonaxial when ambipolar diffusion is
present. Short wavelength unstable wavenumber modes are always
present, even for increasing outward angular velocity profiles. In
a non-accreting system, solid body rotation is an energy extremum.
Differential rotation causes no local linear instabilities in a
purely hydrodynamic gas, and even in the presence of a magnetic
field, ideal MHD only destabilises outwardly decreasing angular
velocity profiles. It is therefore noteworthy that non-ideal MHD
processes open paths to destabilise any differentially rotating
configuration. A likely consequence of the action of the
instability in nonaccreting systems is solid body rotation, which
may explain the velocity profiles of molecular cloud cores
\citep{bg98}. In a disc in which accretion is a possibility, the
nonlinear outcome of the instability will not be solid body
rotation, but pronounced differential rotation and turbulence.

The parameter that measures the relative importance of ambipolar
diffusion is $\Omega/\gamma\rho_i$, the ratio of the angular
rotation velocity to ion-neutral collision rate. Although there
are unstable large wavenumber modes at all values of this ratio,
the growth rate of the instability drops rapidly below $\sim
\Omega$ when this ratio approaches or exceeds unity. The viability
of the instability then depends upon whether the growth time
remains small compared to the expected lifetime of the system in
question. When a background $B_R$ is present, there is the
additional uncertainty of whether the dynamical consequences of a
steady, linear-in-time growth of $B_\phi$ will outpace a more
leisurely growing exponential instability.

A subsequent paper will address the stability of a system in which Ohmic
losses, HEMFs, and ambipolar diffusion are simultaneously present, the
dispersion relation of which is given in the Appendix.

\section*{Acknowledgments}
SAB is pleased to thank the Institut d'Astrophysique de Paris for its
hospitality and generous support. MWK was supported by a Harrison Award
from the University of Virginia.  Support from NASA grants NAG5--13288
and NAG5--9266 is gratefully acknowledged.

\section*{Appendix}
We present here the generalised dispersion relation including all
non-ideal MHD effects: Ohmic diffusion, Hall effect, and ambipolar
diffusion for the case $B_R=0$, as in the text. Although the range
of parameter space where all three terms are comparable is very
small, it is nevertheless useful to have a general dispersion
relation which may be conveniently specialised to a particular
gas. Additionally, the final form of the complete dispersion
relation has a mathematical form which can be written as a natural
generalisation of the (comparatively) simple Hall dispersion
relation of \citet{bate01}. This too is worth noting.

Let us first introduce some preliminary notation. We start with
the generalised resistivity parameters, $\Psi_1$ and $\Psi_2$:
\begin{equation}
\Psi_1 = \eta + {v^2_{AZ}\over \gamma\rho_i}, \quad \Psi_2 = \eta +
{k_Z^2\over k^2}{v^2_{AZ}\over \gamma\rho_i} + {v^2_{A\phi}\over\gamma\rho_i},
\end{equation}
and their respective arithmetic and geometric means, $\Upsilon$
and $\Psi^2$:
\begin{equation}
\Upsilon = {1\over2}(\Psi_1+\Psi_2), \quad \Psi^2=\Psi_1\Psi_2.
\end{equation}

Next, there are two parameters with dimensions of magnetic field
divided by frequency, $\bb{B^+}/\nu,$ and $\bb{B^-}/\nu$:
\begin{equation}
{\bb{B^\pm}\over\nu} = \left(n\over n_e\right) {\bb{B}\over
\omega_{c\mu}} \pm {B_Z B_\phi\over B\gamma\rho_i}\bb{e_R} \;,
\end{equation}
where $\omega_{c\mu}$ is the cyclotron frequency of a particle of
mass $\mu$, the mean mass per neutral particle,
$$
\omega_{c\mu} = {e B\over \mu c},
$$ and $\bb{e_R}$ is the unit vector in the radial direction.

With these definitions, the generalised non-ideal MHD induction
equations may be written:
\begin{equation}\label{totindr}
\left(\sigma + k^2\Psi_1\right)\delta B_{R} +
\frac{k_{Z}B(\bb{k}\bcdot\bb{B^-})}{4\pi\rho\nu}\delta B_{\phi} -
i(\bb{k}\bcdot\bb{B})\delta v_{R} = 0,
\end{equation}\begin{eqnarray}\label{totindphi}
\lefteqn{\left(\sigma + k^2\Psi_2\right)\delta B_{\phi} -
\left[\frac{d\Omega}{d\ln R} +
\frac{k_{Z}B(\bb{k}\bcdot\bb{B^+})}{4\pi\rho\nu}\right]\delta
B_{R}}\nonumber\\*&& \mbox{} +
\frac{k_{R}B(\bb{k}\bcdot\bb{B^+})}{4\pi\rho\nu}\delta B_{Z} -
i(\bb{k}\bcdot\bb{B})\delta v_{\phi} = 0,
\end{eqnarray}\begin{equation}\label{totindz}
\left(\sigma + k^2\Psi_1\right)\delta B_{Z} -
\frac{k_{R}B(\bb{k}\bcdot\bb{B^-})}{4\pi\rho\nu}\delta B_{\phi} -
i(\bb{k}\bcdot\bb{B})\delta v_{Z} = 0.
\end{equation}
The dynamical equations (\ref{dyn1}) and (\ref{dyn2}) remain
unchanged.

The non-ideal MHD dispersion relation may be written in a form
similar to that of the axisymmetric dispersion relation of a disc
in the Hall regime \citep{bate01}:
\begin{eqnarray}
\lefteqn{\sigma^{4} + 2k^{2}\Upsilon\sigma^{3} +
\mathcal{K}_{2}\sigma^{2} +
2k^{2}\Upsilon\left[(\bb{k}\bcdot\bb{v_{\rm A}})^{2} +
\frac{k^{2}_{Z}}{k^{2}}\kappa^{2}\right]\sigma + \mathcal{K}_{0} =
0,}\nonumber\\*
\end{eqnarray}
with
\begin{eqnarray}
\lefteqn{\mathcal{K}_{2} = \frac{k^{2}_{Z}}{k^{2}}\kappa^{2} +
2(\bb{k}\bcdot\bb{v_{\rm A}})^{2} + k^{4}\Psi^{2} +
\frac{k^{2}}{k^{2}_{Z}}\frac{k_ZB(\bb{k}\bcdot\bb{B^{-}})}{8\pi\Omega\rho\nu}}\nonumber\\*&&
\mbox{}\times\left[\frac{k_{Z}\Omega
B(\bb{k}\bcdot\bb{B^+})}{2\pi\rho\nu}+\frac{k^{2}_{Z}}{k^{2}}\frac{d\Omega^2}{d\ln
R}\right],
\end{eqnarray}\begin{eqnarray}\label{k0}
\lefteqn{\mathcal{K}_{0} = \kappa^{2}k^{2}_{Z}k^{2}\Psi^{2} +
\left[(\bb{k}\bcdot\bb{v_{\rm A}})^{2} + \frac{k_{Z}\Omega B
(\bb{k}\bcdot\bb{B^+})}{2\pi\rho\nu} +
\frac{k^{2}_{Z}}{k^{2}}\frac{d\Omega^{2}}{d\ln R}\right]
}\nonumber\\*&& \mbox{}\times\left[(\bb{k}\bcdot\bb{v_{\rm
A}})^{2} + \frac{\kappa^{2}}{4\Omega^{2}}\frac{k_{Z}\Omega B
(\bb{k}\bcdot\bb{B^-})}{2\pi\rho\nu}\right].
\end{eqnarray}
A detailed analysis of this equation is deferred to a later paper.

\bsp\label{lastpage}

\end{document}